\documentclass{frontiersSCNS} 

\usepackage{url,lineno}
\usepackage{epsfig}

\copyrightyear{}
\pubyear{}

\def\journal{Interdisciplinary Physics}
\def\DOI{}
\def\articleType{Research Article}
\def\keyFont{\fontsize{8}{11}\helveticabold }
\def\firstAuthorLast{Danku {et~al.}} 
\def\Authors{Zsuzsa Danku and Ferenc Kun$^{*}$}
\def\Address{Department of Theoretical Physics, University of Debrecen, P.O.Box:
5, Debrecen, H-4010, Hungary}
\def\corrAuthor{Ferenc Kun}
\def\corrAddress{Department of Theoretical Physics, University of Debrecen,
P.O.Box: 5, Debrecen, H-4010, Hungary}
\def\corrEmail{ferenc.kun@science.unideb.hu}

\begin{document}
\onecolumn
\firstpage{1}

\title[Record breaking in creep rupture]{Record breaking bursts in a fiber 
bundle model of creep rupture}
\author[\firstAuthorLast ]{\Authors}
\address{}
\correspondance{}
\extraAuth{}
\topic{Research Topic}

\maketitle
\begin{abstract}

\section{
We investigate the statistics of record breaking events in the time series of 
crackling
bursts in a fiber bundle model of the creep rupture of heterogeneous materials. 
In the model fibers break due to two mechanisms: 
slowly accumulating damage triggers bursts of immediate breakings analogous
to acoustic emissions in experiments. The rupture process accelerates such that 
the size of breaking 
avalanches increases while the waiting time between consecutive events decreases 
towards failure.
Record events are defined as bursts which have a larger size than all previous 
events in the 
time series.
We analyze the statistics of records focusing on the limit of equal load sharing 
(ELS) 
of the model and compare 
the results to the record statistics of sequences of independent identically 
distributed 
random variables. Computer simulations revealed that the number of records grows 
with the logarithm 
of the event number except for the close vicinity of macroscopic failure where 
an exponential dependence is
evidenced. The two regimes can be attributed to the dominance of disorder with 
small burst sizes
and to stress enhancements giving rise efficient triggering of extended bursts, 
respectively.
Both the size of records and the increments between consecutive record events
are characterized by power law distributions with a common exponent 1.33 
significantly different
from the usual ELS burst size exponents of fiber bundles. The distribution of 
waiting times follows
the same behavior, however, with two distinct exponents for low and high loads. 
Studying the 
evolution of records we identify a load dependent characteristic scale of the 
system
which separates slow down and acceleration of record breaking as failure is 
approached.
}

\tiny
 \keyFont{ \section{Keywords:} Fracture, crackling noise, fiber bundle model, 
avalanche,
record breaking statistics}
\end{abstract}

\section{Introduction}

The fracture of heterogeneous materials proceeds in bursts generated by newly
nucleating cracks or by intermittent propagation steps of crack fronts 
\cite{alava_statistical_2006}. 
Measuring acoustic emissions of breaking bursts
the fracture process can be decomposed into a time series of crackling events,
which contains valuable information about the microscopic dynamics of fracture
\cite{deschanel_experimental_2009,salminen_acoustic_2002,rosti_statistics_2010}.
The analysis of crackling time series has usually been focused
on the integrated statistics of events such as the probability distribution 
of the size (energy) and duration of bursts and of the waiting time between 
consecutive events. All these distributions are found to have power law 
functional
form with exponents having a high degree of robustness with respect to 
materials' 
details  
\cite{deschanel_experimental_2009,salminen_acoustic_2002,rosti_statistics_2010}.

In applications materials are often subject to constant sub-critical loads
which may lead to failure in a finite time. Such creep rupture processes 
are typical for components of engineering constructions and they also play
a crucial role in the emergence of natural catastrophes such as land slides, 
stone and
snow avalanches, and earthquakes 
\cite{nechad_creep_2005,nechad_andrade_2005,petri_experimental_1994,
garcimar_statistical_1997,main_damage_2000,
santucci_sub-critical_2004,deschanel_statistical_2006,santucci_subcritical_2006,
kovacs_critical_2008,PhysRevLett.110.165506}. 
It is of high practical importance to understand 
how the creeping system approaches macroscopic failure using the data of 
acoustic monitoring 
\cite{garcimar_statistical_1997,nechad_creep_2005,
kovacs_critical_2008,deschanel_experimental_2009}. 
Computer simulations of discrete stochastic models of creep rupture are 
indispensable to 
analyze how the time series of crackling events evolves and to identify possible 
signatures
of the imminent catastrophe. A unique feature of this evolution is that the 
rupture process 
is highly non-stationary, i.e.\ approaching failure larger and larger bursts are 
triggered while
the process accelerates indicated by the decreasing waiting time between 
consecutive events
\cite{garcimar_statistical_1997,nechad_creep_2005}.
On the macro-scale, the strain rate has been found to exhibit 
 time-to-failure power law behavior 
\cite{nechad_creep_2005,nechad_andrade_2005,deschanel_statistical_2006,
deschanel_experimental_2009} 
which is accompanied by the emergence of an Omori-type 
acceleration of the rate bursts on the micro-scale \cite{danku_creep_2013}.

In the present paper we investigate
the evolution of the crackling time series of creep rupture by analyzing the 
statistics
of record breaking (RB) bursts in a fiber bundle model (FBM) of creep failure.
Records are bursts which have the largest size since the beginning of the time 
series,
hence, their behavior involves  extreme value statistics
\cite{galambos_asymptotic_1978,record_book_1}. Motivated mainly by climate 
research \cite{PhysRevE.74.061114} 
and by the investigation
of earthquake time series 
\cite{npg-17-169-2010,PhysRevE.77.066104,PhysRevE.87.052811} interesting 
analytical results have recently been obtained for the record breaking 
statistics of sequences 
of independent identically distributed (IID) random variables. 
The statistics of records has proven useful 
to identify trends in time series of measurements and to infer correlations of 
events 
\cite{PhysRevE.87.052811,npg-17-169-2010}.
Focusing on the limit of equal load sharing of our FBM we demonstrate that the 
record breaking statistics
of crackling events provides novel insight into rupture phenomena. Comparing the 
outcomes 
of large scale computer simulations to the corresponding IID results on records 
we can identify regimes of the failure process dominated by the disorder of 
materials 
and by the enhanced triggering of breaking avalanches towards failure. 
The size distribution of records proved
to have a power law form with a novel exponent of equal load sharing FBMs. 
Simulations revealed the emergence of a load dependent characteristic record 
rank
which separates slow down and acceleration of record breaking when approaching 
failure.

\section{Material \& Methods}
To investigate the time series of breaking bursts
we use a generic fiber bundle model (FBM) of creep rupture introduced recently
\cite{kun_fatigue_2007,1742-5468-2009-01-P01021,kun_universality_2008,
PhysRevE.85.016116,
danku_creep_2013,danku_PhysRevLett.111.084302}.
We briefly summarize the main ingredients of the model construction emphasizing
aspects most relevant for the present study.

\subsection{Fiber bundle model with two breaking mechanisms}
In the framework of the model the sample is discretized in terms of a bundle 
of parallel fibers which have a brittle response with identical Young modulus 
$E$. The bundle is 
subject to a constant external load $\sigma_0$ below the fracture strength
$\sigma_c$ of the system parallel to the fibers' direction. 
Fibers are assumed to break due to two physical mechanisms:
immediate breaking occurs when the local load $\sigma_i$ on fibers exceeds 
their fracture strength 
$\sigma_{th}^{i}$, $i=1,\ldots , N$. Under a sub-critical load $\sigma_0 < 
\sigma_c$ this 
breaking mechanism would lead to a partially
failed configuration with an infinite lifetime.
Time dependence is introduced in such a way that those fibers, which remained
intact, undergo an aging process accumulating damage $c(t)$. 
The damage mechanism represents the environmentally induced slowly developing
aging of materials such as corrosion cracking and thermally or chemically 
activated
degradation 
\cite{kun_fatigue_2007,1742-5468-2009-01-P01021,kun_universality_2008,
PhysRevE.85.016116,danku_creep_2013}
similar to damage dynamics based models of rock fracture developed for instance 
in \cite{GJI:GJI1884}. 
The rate of damage accumulation $\Delta c_i$ is assumed 
to have a power law dependence on the local load 
\begin{eqnarray}
\Delta c_i = a\sigma_i^{\gamma}\Delta t,
\label{eq:damlaw}
\end{eqnarray}
where $a$ is a constant and the exponent $\gamma$ controls the time scale of
the aging process with $0\leq \gamma < +\infty$. 
The total amount of damage $c_i(t)$ accumulated up to 
time $t$ is obtained
by integrating over the entire loading history of fibers 
$c_i(t)=a\int_0^t\sigma_i(t')^{\gamma}dt'$. 
Fibers can tolerate only a finite amount of 
damage so that when $c_i(t)$ exceeds the local damage threshold $c^i_{th}$ the
fiber breaks.
Each breaking event is followed by a redistribution of load over the remaining
intact fibers. Two limiting cases of load sharing are usually considered in 
FBMs: under equal load
sharing (ELS) conditions all the intact fibers keep the same amount of load so 
that the load on 
a single fiber is $\sigma_i=N\sigma_0/(N-i)$ after the breaking of $i$ fibers. 
ELS realizes the mean field limit of FBMs where no stress heterogeneity can 
arise. 
In the opposite limit of localized load sharing (LLS) the load of broken fibers 
is equally 
redistributed solely over their intact nearest neighbors in the bundle. LLS 
leads to a high 
stress concentration around
failed regions and it gives rise to the emergence of spatial correlations 
between consecutive 
breaking events. 
In the present study we consider only the ELS case
where the homogeneous stress field hinders spatial correlation. 
In the model the quenched heterogeneity of materials is represented by 
the randomness of breaking thresholds $\sigma^{th}_i, c^{th}_i$, $i=1,\ldots , 
N$. 
For simplicity, for both threshold values we assume uniform distributions 
between zero and one. 
Since under ELS the value of the exponent $\gamma$ only controls the time scale 
of creep 
\cite{danku_creep_2013} the damage parameters were fixed to $\gamma=1$ and 
$a=1$.

When the load is put on the bundle, first some week fibers break immediately 
which may generate
further breakings until a stable configuration is reached where all remaining 
intact fibers can 
sustain the elevated load. The time evolution of the creeping system starts from 
this partially 
failed configuration \cite{kun_fatigue_2007,1742-5468-2009-01-P01021}. The 
present setup of the model 
implies that the critical stress where immediate catastrophic failure occurs is 
equal to its static value 
$\sigma_c=0.25$ 
\cite{kun_fatigue_2007,1742-5468-2009-01-P01021,kun_universality_2008}. In the 
presentation of the results the constant external load level will be 
characterized
by the ratio $\sigma_s=\sigma_0/\sigma_c$ which can take values in the range $0 
< \sigma_s\leq 1$.

\subsection{Bursts driven by damage sequences}

The separation of time
scales of the slow damage process and of immediate breaking leads to a highly 
complex 
time evolution in agreement with experiments 
\cite{1742-5468-2009-01-P01021,kun_universality_2008,PhysRevE.85.016116,
danku_creep_2013,danku_PhysRevLett.111.084302}: 
starting from the 
initial configuration damaging fibers break slowly one-by-one 
gradually increasing the load
on the remaining intact fibers. After a certain number $\Delta_d$ of damage 
breakings the load increment
becomes sufficient to induce the immediate breaking of a fiber which in turn 
triggers
an entire burst of immediate breakings. As a consequence, the time evolution of 
creep rupture
occurs as a series of bursts corresponding to the nucleation and propagation of 
cracks,
separated by silent periods of slow damaging. The size of bursts $\Delta$ is 
defined as the number 
of fibers breaking in avalanches. The number of fibers breaking in a damage 
sequence
and its duration determine the length of the damage sequence $\Delta_d$ and the 
physical waiting time $T$ between consecutive events, respectively. 

\begin{figure}
\begin{center}
\epsfig{bbllx=5,bblly=5,bburx=625,bbury=300,file=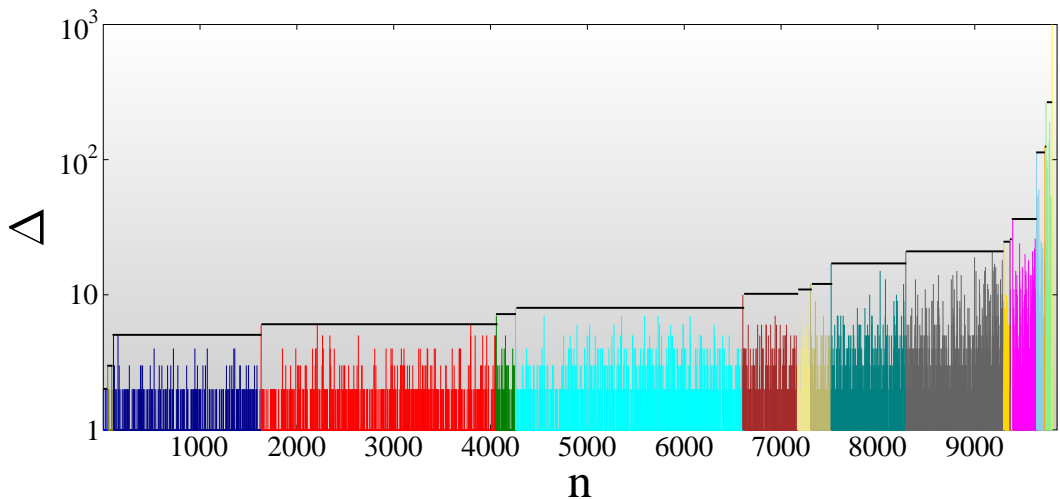, 
width=14.0cm}
\end{center}
  \caption{Time series of bursts in a mean field (ELS) simulation of the FBM 
of creep. For demonstration purposes a relatively small system $N=10^5$ is 
presented
at the load $\sigma_s= 0.001$ giving rise to $9793$ avalanches. In order to have 
a clear
view on the structure of the time series, sub-sequences starting at record 
breaking events 
are highlighted by different colors. The horizontal lines serve to demonstrate 
the increments
between consecutive record bursts.
  \label{fig:rbsequence_demo}
}
\end{figure}

Figure \ref{fig:rbsequence_demo} shows a representative example of the time 
series of 
bursts as the system evolves towards failure. In order to have a clear view on 
the details 
of the sequence of events we intentionally used a relatively small system of 
$N=10^5$ fibers
subject to the load $\sigma_s=0.001$ which gave rise to 9793 bursts. 
The size $\Delta$ of bursts is shown in the figure as a function of the discrete 
time, 
i.e.\ order number or natural time $n=1,2,\ldots$
so that no information is presented about the physical waiting time $T$ elapsed 
between the bursts.
Strong fluctuations of the burst size $\Delta$ can be observed 
which is caused by the quenched disorder of fiber strength. The time series has 
a non-stationary behavior
which is indicated by the increasing average size of bursts. 
Of course, the physical waiting time $T$ decreases between consecutive events, 
however, this information
is not visible in this representation. (The analysis of waiting times of the 
model can be found in 
\cite{kun_universality_2008,1742-5468-2009-01-P01021,danku_creep_2013}.)

A record of the time series is a burst which has a size $\Delta_{r}$ larger than 
any previous events.
Consecutive records are identified by their increasing rank $k$ as $k=1,2,3, 
\ldots$ which occurred
as the $n_k$th burst of the complete time series. The first burst $n=1$ is by 
definition considered
to be a record of rank $k=1$ with $n_1=1$. In the example of Fig.\ 
\ref{fig:rbsequence_demo} all
together 18 record breaking events are identified $\Delta_r^k$  ($k=1,\ldots , 
18$),
which are highlighted by using different colors for the consecutive smaller 
events. 
It can be observed that RB events form a sub-sequence of bursts 
with monotonically increasing size, however, both the record size and the number 
of avalanches 
between two consecutive records exhibit strong fluctuations.
In order to characterize these features we introduce the size increment  
$\delta_k$ 
and the waiting time $m_k$ between two records with the definitions 
\begin{eqnarray}
\delta_k=\Delta_r^{k+1}-\Delta_r^k, \ \ \ \ \ \ \ \ \ \ \ \mbox{and} \ \ \ \ \ \ 
\ \ \ \ \ m_k=n_{k+1}-n_k, 
\label{eq:wait}
\end{eqnarray}
respectively. Note that the catastrophic burst which breaks all remaining
intact fibers and destroys the bundle is not included in the time series so
that the last burst in the bundle may not be an RB event.

\section{Results}
To investigate the occurrence of record breaking events during the rupture 
process computer simulations were 
carried out for a system of size $N=10^7$ fibers averaging over $10^4$ 
realizations
of the threshold disorders at each load value $\sigma_s$. The external load 
$\sigma_s$
was varied over a broad range $0.001 \leq \sigma_s \leq 0.9$, where the limits 
were set to have a sufficient
number of bursts in the time series.
\begin{figure}
\begin{center}
\epsfig{bbllx=30,bblly=470,bburx=1060,bbury=760,
file=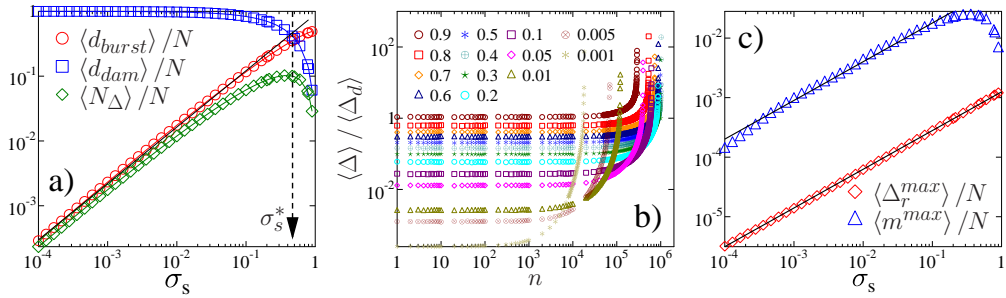, width=13.5cm}
\end{center}
  \caption{$(a)$ Fraction of fibers breaking in bursts $\left<d_{burst}\right>$ 
and 
due to damage $\left<d_{dam}\right>$ as function of load $\sigma_s$ together 
with 
the number of bursts $\left<N_{\Delta}\right>$. The quantity 
$\left<d_{burst}\right>$
was calculated as the sum of the size of all bursts that occurred up to failure.
The straight line has slope 0.87, furthermore, the vertical dashed line 
indicates the position of the
characteristic load $\sigma_s^* = 0.48$. 
$(b)$ Ratio of the average size of bursts $\left<\Delta\right>$ and of damage 
sequences $\left<\Delta_d\right>$
as a function of the order number $n$ of bursts for several load values.
$(c)$ Average size of the largest record breaking event
$\left<\Delta_r^{max}\right>$ and the average value of the largest waiting time 
$\left<m^{max}\right>$ between 
consecutive records as function of load. All quantities are normalized by the 
number of fibers $N$
in the bundle. The upper and lower straight lines have slope 0.625 and 0.65, 
respectively.        
  \label{fig:aver_burst}
}
\end{figure}
We identify all record breaking avalanches $\Delta_r^k$ that occur up to 
macroscopic failure of the bundle
and carry out a detailed analysis of their statistics.

\subsection{Number of record breaking avalanches}
An important feature of our system is that under low load values most of the 
fibers break 
due to slow damaging since the resulting load increments are too small to 
trigger extended
bursts of immediate breakings. Figure \ref{fig:aver_burst}$(a)$ shows that the 
fraction
of fibers breaking in avalanches $\left<d_{burst}\right>$ and due to damage 
$\left<d_{dam}\right>$
are monotonically increasing and decreasing functions of the external load 
$\sigma_s$, respectively.
The two curves intersect each other at the characteristic load 
$\sigma_s^*\approx 0.48$, which
coincides with the position of the maximum of the average number of avalanches 
$\left<N_{\Delta}\right>$. 
The result demonstrates that for $\sigma_s < \sigma_s^*$ damage dominates the 
failure process, while
the vicinity of catastrophic failure $\sigma_s > \sigma_s^*$ is controlled by 
the bursting activity with 
large burst sizes $\Delta$ and a decreasing number $N_{\Delta}$ of bursts 
(see also \cite{kun_universality_2008}). 

\begin{figure}
\begin{center}
\epsfig{bbllx=10,bblly=170,bburx=730,bbury=770,
file=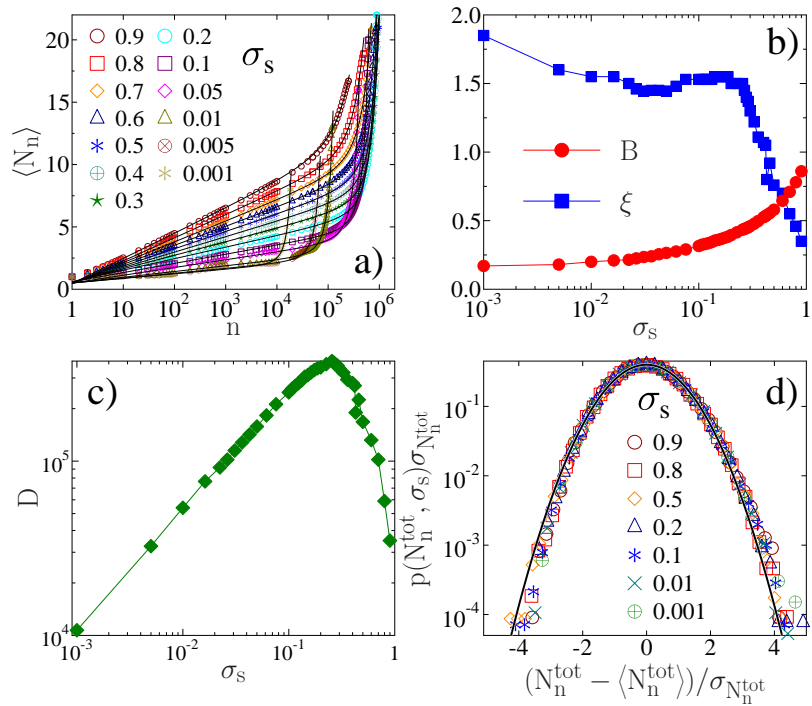, width=11.0cm}
\end{center}
  \caption{$(a)$ Average number of records $\left<N_n\right>$ as a function of 
the total
  number of avalanches $n$ for several load values. For low record numbers 
logarithmic
  dependence is evidenced, which is followed by a faster exponential increase. 
  The continuous black lines represent fits with Eq.\ (\ref{eq:log}).
  $(b,c)$ Fitting parameters
  $B$, $\xi$, and $D$ of Eq.\ (\ref{eq:log}) as function of the external load 
$\sigma_s$.
  $(d)$ Probability distribution of the total number of records $N_n^{tot}$ 
occurred up to failure
  for several load values. The distributions obtained at different loads
  are rescaled with the corresponding average and standard deviation. The 
continuous line 
  represents the standard Gaussian  Eq.\ (\ref{eq:gaussian}). 
  \label{fig:rbnumber}
}
\end{figure}

Figure \ref{fig:rbnumber}$(a)$ presents the average number of records 
$\left<N_n\right>$ 
that occurred until $n$ avalanches have been generated  in the time series at 
several load values. 
Due to the breaking dynamics described above at low loads avalanches remain 
small typically 
comprising 
a few breaking fibers, and hence, new records mainly occur close to macroscopic 
failure. 
For instance, at loads $\sigma_s \leq 0.01$ 
the number of records has very low values $\left<N_n\right><3$ 
up to large $n$ followed by a fast increase in the vicinity of the failure 
point. 
As $\sigma_s$ increases the qualitative form of the $\left<N_n\right>$ curves 
remains the same, 
they just shift to higher record numbers due to the more intensive triggering of 
larger bursts
at high loads. The most 
important feature of the record number $\left<N_n\right>$ is that it has a 
logarithmic dependence 
on $n$ over a broad range, which is in agreement with the analytic prediction of 
the logarithmic dependence
of record numbers of IIDs on the event number $n$ 
\cite{record_book_1,PhysRevE.87.052811}.
The result implies that except for the close vicinity of macroscopic failure 
disorder dominates
the process of creep rupture, and the occurrence of breaking avalanches can be 
well approximated 
as a stochastic process of IIDs. However, close to failure the increase of the 
stress on single fibers
results in enhanced triggering which in turn gives rise to a sudden increase of 
records.
The complete curves of $\left<N_n\right>(n)$ can be characterized by the 
functional form
\begin{eqnarray}
\left<N_n\right>=A+B\ln{n}+C\exp{\left[(n/D)^{\xi}\right]},
\label{eq:log} 
\end{eqnarray}
where the exponential term describes the rapid generation of high rank records
close to catastrophic failure. 
It can be observed in Fig.\ \ref{fig:rbnumber}$(a)$ that 
Eq.\ (\ref{eq:log}) provides an excellent fit of the numerical data over the 
complete load range considered.
The additive parameter $A$ has the same value $A=0.38$ for all loads, while the 
multiplication
factor $B$ increases from 0.16 to 0.87 in the load range considered (see Fig.\ 
\ref{fig:rbnumber}$(b)$).
The shape of the $\left<N_n\right>$ curves in the accelerating regime is 
controlled by the exponent $\xi$
which is nearly constant $\xi\approx 1.5$ for loads below $\sigma_s^*$ then it 
decreases to 0.33 
as $\sigma_s$ increases in Fig.\ \ref{fig:rbnumber}$(b)$. 
The scale parameter $D$ mainly sets the transition point between the two regimes 
dominated 
by disorder and by stress enhancement. It can be seen in Fig.\ \ref{fig:rbnumber}$(c)$ 
that $D$ increases
with the external load and has a maximum in the vicinity of $\sigma_s^*$, which 
is consistent
with the behaviour of the total number of bursts $\left<N_{\Delta}\right>$
in Fig.\ \ref{fig:aver_burst}$(a)$ reflecting the overall dominance of damage 
and bursts
in the failure process on the two sides of $\sigma_s^*$.
To support the emergence of a characteristic event number $D$ separating 
different regimes of
bursting activity, Fig.\ \ref{fig:aver_burst}$(b)$ presents the ratio of the 
average size 
of bursts $\left<\Delta\right>(n)$ and the average size of the damage sequence 
$\left<\Delta_d\right>(n)$ that initiated the $n$th event. At the beginning of 
the process 
a large number of fibers must break randomly one-by-one in damage sequences 
until a small burst
of size 1-2 fibers is initiated giving rise to 
$\left<\Delta\right>/\left<\Delta_d\right> < 1$.
Comparing to Fig.\ \ref{fig:rbnumber}$(a)$ the onset of the rapid exponential increase 
of record breakings corresponds to the event number from where extended bursts are triggered by 
shorter damage sequences due to the effect of stress enhancements. 
As the consequence of the increasing burst size and shortening damage sequences it becomes more 
and more probable that a record gets broken after a fewer bursts leading to the exponential increase
of the record number. 

 Figure \ref{fig:rbnumber}$(b)$ shows that the probability 
distribution $p(N_n^{tot},\sigma_s)$  of the total number of records accumulated 
up to failure has a 
Gaussian form.  The 
distributions $p(N_n^{tot},\sigma_s)$ obtained at different loads were rescaled 
by 
the corresponding average $\left<N_n^{tot}\right>$ and standard 
deviation $\sigma_{N_n^{tot}}$ of the total number of records 
which results in a high quality collapse in the figure. 
The scaling function has a good agreement with the standard Gaussian 
\begin{eqnarray}
p(x) = \frac{1}{\sqrt{2\pi}}\exp{\left(-x^2/2\right)},
\label{eq:gaussian}
\end{eqnarray}
which has been predicted for RB sequences of IIDs 
\cite{record_book_1,PhysRevE.87.052811}. 
The Gaussian functional
form prevails in spite of the complex effect of the acceleration of the 
occurrence of records 
towards failure. Deviations can be observed at the tails of the distribution 
which may indicate
a slight right-handed asymmetry.   

\subsection{Statistics of record sizes and waiting times}
Recently, it has been shown that in our model the size distribution 
of avalanches $p(\Delta)$ accumulating all crackling events 
up to failure during the creep process follows a power law distribution 
\cite{kun_universality_2008,1742-5468-2009-01-P01021}
\begin{eqnarray}
p(\Delta) \sim \Delta^{-\tau},
\end{eqnarray}
where the value of the exponent $\tau$ coincides with the usual mean field 
exponent of FBMs
$\tau=5/2$ \cite{kloster_burst_1997}. 
As the external load approaches the critical value $\sigma_s\to 1$ a crossover 
is obtained
to a lower exponent $\tau=3/2$ \cite{kun_universality_2008}
in agreement with the mean field prediction of simple FBMs
subject to a quasi-statically increasing load 
\cite{pradhan_failure_2010,pradhan_crossover_2005-1,raischel_local_2006}. 

\begin{figure}
\begin{center}
\epsfig{bbllx=20,bblly=120,bburx=730,bbury=750,
file=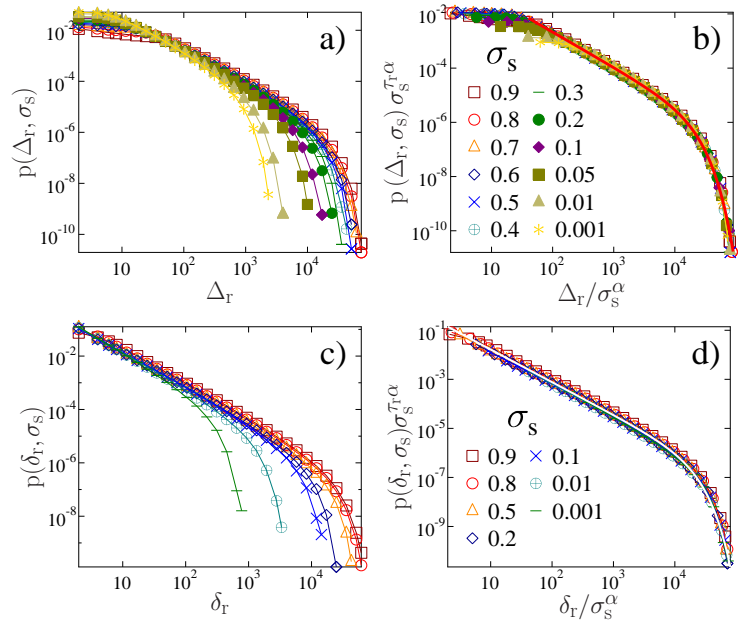, width=9.5cm}
\end{center}
  \caption{ Size distribution of record breaking bursts $(a)$ and their 
increments $(c)$
  for several load values. $(b)$ and $(d)$ demonstrate that rescaling the 
distributions with
  the external load high quality collapse of the curves can be achieved. 
  The continuous lines in $(b)$ and $(d)$ represent fits with Eq.\ 
(\ref{eq:rec_size}).
  \label{fig:rbsize_inc}
}
\end{figure}

Figure \ref{fig:rbsize_inc}$(a)$ presents the size distribution 
$p(\Delta_r,\sigma_s)$ 
of record breaking bursts accumulating all records up to failure at different 
load values. 
The distributions can be well described by a power law followed by a finite size 
cutoff
of exponential form. Since at higher loads larger avalanches are triggered 
the cutoff of the distributions shifts to higher values but the functional form 
remains 
the same for all loads.
Figure \ref{fig:rbsize_inc}$(b)$ illustrates that rescaling the two axis 
in Fig.\ \ref{fig:rbsize_inc}$(a)$ with appropriate powers of the external load 
a high quality data collapse can be achieved. The result implies that record 
size distributions
obtained at different loads obey the scaling structure
\begin{eqnarray}
p(\Delta_r,\sigma_s) = \Delta_r^{-\tau_r}\phi(\Delta_r/\sigma_s^{\alpha}),
\label{eq:rec_size}
\end{eqnarray}
where the exponent $\alpha$ controls the load dependence of the cutoff record 
\begin{eqnarray}
\left<\Delta_r^{max}\right>\sim \sigma_s^{\alpha}. 
\end{eqnarray}
The power law dependence of the cutoff record size $\left<\Delta_r^{max}\right>$ 
is also
confirmed by Fig.\ \ref{fig:aver_burst}$(c)$, where the largest records 
$\Delta_r^{max}$ 
of single simulations
were directly averaged. Best fit is obtained with the exponent $\alpha=0.65$ 
which provides
also the best collapse in Fig.\ \ref{fig:rbsize_inc}$(b)$.
Note that due to the normalization of the distributions
the scaling exponent used for the transformation along the vertical axis has to 
be equal to the product of
$\tau_r$ and $\alpha$ which is explicitly indicated in Fig.\ 
\ref{fig:rbsize_inc}$(b)$.
The size distribution exponent of record bursts 
$\tau_r=1.33\pm 0.03$ proved to be significantly smaller than the usual mean 
field burst size exponents of FBMs 
\cite{kun_universality_2008,1742-5468-2009-01-P01021,kloster_burst_1997,
pradhan_crossover_2005-1,pradhan_failure_2010}, which 
is the consequence of the relatively high frequency of 
large burst sizes in the 
RB sequence with respect to the complete time series. The short flat region for 
the smallest bursts
occurs due to down sampling of small sized events in the RB sequence.
To fit the scaling function in Fig.\ \ref{fig:rbsize_inc}$(b)$ the cutoff 
function $\phi(x)$  was 
assumed to have a stretched exponential form $\phi(x)\sim 
\exp(-(x/x_0)^{\beta})$ with $\beta=0.6$5. 
It is important to emphasize that the distribution $p(\Delta_r,\sigma_s)$ has a 
homogeneous evolution 
with increasing external load, i.e.\ the exponent $\tau_r$ remains constant when 
the critical load is approached 
$\sigma_s\to 1$.

For the advancement of the RB sequence of bursts during the evolution of the 
rupture process 
the size increments $\delta_k=\Delta_r^{k+1}-\Delta_r^{k}$ between consecutive 
records carry also interesting
information. Figure \ref{fig:rbsize_inc}$(c)$ presents that the distribution 
$p(\delta_r,\sigma_s)$ 
of increments $\delta_r$ has a qualitatively similar behavior to record sizes 
$p(\Delta_r,\sigma_s)$,
i.e.\ power laws are obtained followed by an exponential cutoff as 
described by Eq.\ (\ref{eq:rec_size}).  Careful scaling analysis in
Fig.\ \ref{fig:rbsize_inc}$(d)$ shows that both the value of the exponent 
$\tau_r$ of the power law regime
and the scaling exponent $\alpha$ of $\sigma_s$ have the same values for the two 
distributions, the 
only difference is that the cutoffs of increments are smaller than the one of 
record sizes.

\begin{figure}
\begin{center}
\epsfig{bbllx=20,bblly=460,bburx=1110,bbury=760,
file=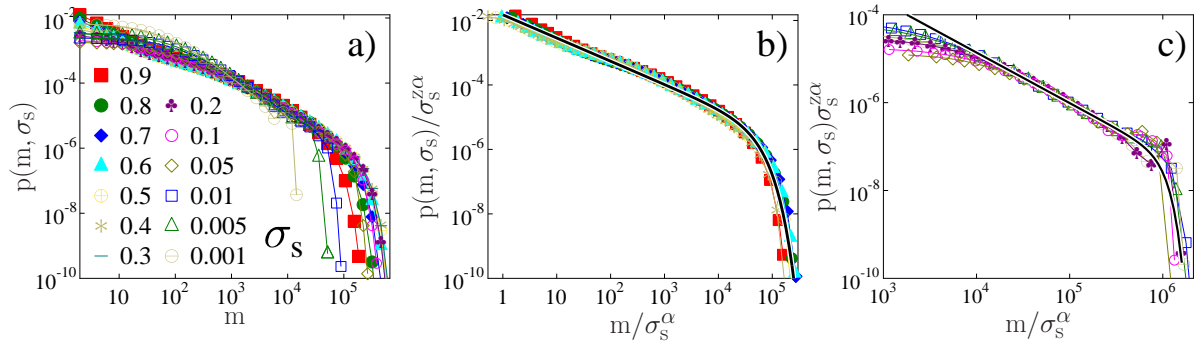, width=16.0cm}
\end{center}
  \caption{$(a)$ Probability distribution of waiting times $p(m,\sigma_s)$ 
between consecutive records for
several values of the external load $\sigma_s$. In $(b)$ and $(c)$ the same 
distributions are presented 
after rescaling the two axis with powers of the external load $\sigma_s$ 
separately above and 
below $\sigma^*_s$, respectively.
The good quality data collapse demonstrates the validity of the scaling 
structure Eq.\ (\ref{eq:scal_wait}).
  \label{fig:rbwaiting_time}
}
\end{figure}

After a record occurred as the $n_{k}$th avalanche of the time series it gets 
broken after a certain number of events by the $n_{k+1}$th avalanche. The 
waiting time $m_k$ 
between RB avalanches defined by Eq.\ (\ref{eq:wait}) is an important 
characteristic quantity of the RB sequence of bursts. 
Figure \ref{fig:rbwaiting_time}$(a)$ presents the accumulated statistics of 
waiting times $m$ 
considering all records $k$ in the RB sequence for several load values. 
Based on the statistics of extremes 
\cite{galambos_asymptotic_1978}, for IIDs a power law behavior is expected 
$p(m) \sim m^{-z}$ with the exponent $z_{IID}=1$ \cite{record_book_1}. 
It can be observed in the figure that our results are generally consistent 
with the IID prediction, however, distributions $p(m,\sigma_s)$ obtained below 
and above $\sigma_s^*$ 
form two groups of different power law exponents. 
Figures \ref{fig:rbwaiting_time}$(b)$ and $(c)$ demonstrate that rescaling $m$ 
and the distributions 
$p(m,\sigma_s)$ with appropriate powers of the external load $\sigma_s$, good 
quality data collapse 
can be achieved in both load regimes. The scaling structure is similar to Eq.\ 
(\ref{eq:rec_size}) 
\begin{eqnarray}
p(m) =m^{-z}\psi(m/\sigma_s^{\alpha}), 
\label{eq:scal_wait}
\end{eqnarray}
however, both exponents $z$ and $\alpha$ proved to have different values below 
and above the characteristic 
load $\sigma^*_s$. In Figures \ref{fig:rbwaiting_time}$(b)$ and $(c)$ best 
collapse is obtained with 
the exponents $z=0.72$, $\alpha=-1.45$ ($\sigma^*_s < \sigma_s$)
and $z=1.15$, $\alpha=0.625$ ($\sigma^*_s > \sigma_s$). 
The value of $z$ reflects an interesting aspect of the dynamics of the creep 
process:
the low value of $z<z_{IID}$ at high loads shows that long waiting times more 
frequently occur
than for IID. The reason is that due to the large bursts triggered under high 
external loads 
it takes longer for the system to break a record. At low loads the waiting time 
exponent $z$ is slightly larger than the IID prediction implying an elevated 
frequency
of short waiting times with respect to IIDs.
Note that $\alpha$ has different signs in the two regimes
corresponding to the increasing and decreasing behavior of the cutoff of the 
distribution with increasing 
load. An independent test of the load dependence of the cutoff waiting time 
$\left<m^{max}\right>$ 
is shown in Fig.\ \ref{fig:aver_burst}$(c)$, where $m^{max}$ was directly 
averaged over the simulations.
A maximum is obtained at $\sigma^*_s$ in agreement with the scaling behavior in 
Figs.\ 
\ref{fig:rbwaiting_time}$(b)$ and $(c)$. The flattening of the distributions 
$p(m,\sigma_s)$ for low
$m$ values and the small bump close to the cutoff observed  for low loads 
in Fig.\ \ref{fig:rbwaiting_time}$(c)$ are caused by the finite system size and 
by the 
distinct distribution of the time needed to break the first record, 
respectively. 

\subsection{Evolution of the sequence of records}
\begin{figure}
\begin{center}
\epsfig{bbllx=30,bblly=180,bburx=710,bbury=780,file=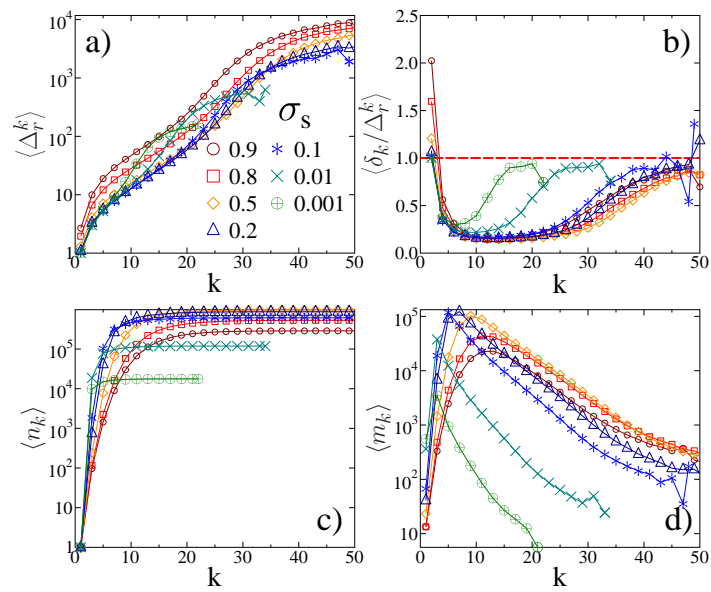, width=9.0cm}
\end{center}
  \caption{Characteristic quantities of single records: the average value of the 
size 
$\left<\Delta_r^k\right>$ $(a)$, and relative increments 
$\left<\delta_k/\Delta_r^{k}\right>$ 
of record breaking events $(b)$. In  $(b)$ the horizontal dashed line highlights 
the 
limit value $1$. Time of occurrence $\left<n_k\right>$ $(c)$ 
and waiting time $\left<m_k \right>$ between consecutive records $(d)$.
  \label{fig:k_averages}
}
\end{figure}

The creeping system approaches macroscopic failure through an accelerating 
sequence of bursts of increasing size.
In order to understand how record breaking events occur during this evolution 
 we evaluated average values of the characteristic quantities 
of single records as a function of their rank $k$. 
In Fig.\ \ref{fig:k_averages}$(a)$ the average record size 
$\left<\Delta_r^k\right>$ 
has the same generic form for all loads, 
i.e.\ a nearly exponential increase is obtained with a slight minimum of the 
derivative of the curves for
intermediate ranks. Note that for a given value of $k$
a record can have a higher value at low loads than at higher ones, e.g.\ at 
$k=15$ the corresponding
record burst is significantly larger at $\sigma_s=0.001$ than at $\sigma_s=0.1$ 
in spite of the two orders
of magnitude higher external load in the second case. The reason is that for 
avalanche triggering the load on single fibers 
is the most relevant quantity, hence, at the generation of the $k$th record the 
creeping system
can be closer to catastrophic failure at lower loads than at higher ones, which 
implies larger burst sizes.
In order to gain information about the rate of increase of records, we 
determined the relative increments 
of consecutive RB events defined as the ratio $\delta_k/\Delta_r^{k}$. Figure 
\ref{fig:k_averages}$(b)$ shows that 
at the beginning of the rupture process the average 
$\left<\delta_k/\Delta_r^{k}\right>$ 
starts from a high value simply because the first record has size $\Delta_r^1=1$ 
and it gets typically broken
by a burst of size $\Delta_r^2=2$ or $\Delta_r^2=3$. Then the relative increment 
rapidly decreases to the vicinity of $0.25$. 
The remarkable result is that for records of the highest rank 
$\left<\delta_k/\Delta_r^{k}\right>$  tends to 1,
which implies that as the system approaches macroscopic failure record breakings 
occur 
by nearly doubling the size of the previous record. Hence, not only the record 
sizes but also
the increments form a monotonically increasing sequence when approaching 
failure.

As the external load increases larger avalanches are triggered in the system and 
they occur
with an increasing rate \cite{danku_creep_2013}. Based on this general tendency 
it could be expected that the average time $\left<n_k \right>$ where the $k$th 
record appears
is a decreasing function of the external load $\sigma_s$. Figure 
\ref{fig:k_averages}$(c)$ 
shows that for
low load values just the opposite happens: records of a given rank $k$ occur 
later after
a larger number of avalanches. The reason is that as the load increases
early records get larger, and hence, they are more difficult to overcome. The 
situation changes
at the characteristic load $\sigma_s^*$ so that in the range $\sigma>\sigma_s^*$
the average record time $\left<n_k \right>$ decreases with increasing load.

The time series presented in Fig.\ \ref{fig:rbsequence_demo} does not contain 
information
about the physical time elapsed between events. However, the acceleration of the 
system can still
be inferred from the natural time $n$, because the number of bursts between RB 
events
decreases when approaching catastrophic failure.
Due to this acceleration the average waiting time $\left<m_k \right>$
is expected to decrease between consecutive records. 
It is interesting to note that for higher loads the decreasing branch
of $\left<m_k \right>$ is preceded by a rapidly increasing regime indicating the 
slow down
of record breaking before acceleration sets on. The maximum of $\left<m_k 
\right>$
already develops for the lowest loads with the position $k^*=2$ and it gradually 
shifts 
to $k^*=13$ for the highest ones.
Comparing Figs.\ \ref{fig:k_averages}$(b)$ and $(d)$
it can be seen that the record rank $k^*$ of the maximum of $\left<m_k \right>$ 
falls close to the position of the minimum of the relative increments 
$\left<\delta_k/\Delta_r^{k}\right>$. The result demonstrates the emergence of a 
characteristic
time scale $n_{k^*}$ of the system which separates slow down and acceleration
of the dynamics of record breaking. This is also in agreement with the behavior 
of record numbers
in Fig.\ \ref{fig:rbnumber}$(a)$ so that $n_{k^*}$ approximately coincides with
the onset time of the exponential increase of $\left<N_n\right>$.

\section{Discussion}
We investigated the statistics of records in a sequence of crackling avalanches 
which occur 
during the creep rupture of heterogeneous materials. Synthetic sequences of 
bursts were
generated by computer simulations of a realistic fiber bundle model where slowly 
developing 
damage triggers avalanches of immediate breaking of fibers. The bundle is 
subject 
to a constant external load
below the fracture strength of the system. We analyzed
the mean field limit of the model where all fibers keep the same load so that no 
spatial 
correlation develops between local failure events.

Record events are defined solely based on the burst size, i.e.\ a record is a 
bursts whose size is
larger than that of all previous bursts. This way a monotonically increasing 
sub-sequence 
of crackling events is identified. Computer simulations revealed that during the 
evolution 
of the rupture process the average number of records 
increases logarithmically with the number of avalanches except for the close 
vicinity of macroscopic failure where an exponential form is evidenced. 
Additionally, the total number of records obtained up to failure at different 
load values 
has a Gaussian distribution. These findings are in agreement with the robust 
analytic predictions
on the RB statistics of sequences of independent identically distributed random 
variables
which shows that the beginning of the creep process is mainly controlled by the 
quenched disorder of the system. 
The enhanced triggering close to failure due to the rapidly increasing load on 
single
fibers is responsible for the exponential acceleration of record numbers.

The size of records proved to have a power law distribution 
with an exponent $1.33$ significantly lower than the usual mean field exponents 
$5/2$ and $3/2$ of the burst sizes of FBMs. 
The size increments between consecutive records are found to have the same 
scaling structure and
the value of exponents as the record size. To prove the independence of the 
exponents on the external
load a careful data collapse analysis was performed. 
The probability distribution of waiting times between consecutive events has 
also a power law
functional form, however, with different exponents below and above the 
characteristic load
$\sigma_s^*$.
In order to characterize the evolution of the sequence of records we studied the 
average value of the 
relative increment and of the waiting time between consecutive events as a 
function of
the record rank. Both quantities show the emergence of a load dependent 
characteristic
scale in the system: at the beginning a slow down of record breaking occurs with 
an
increasing waiting time and decreasing relative increment. Beyond a 
characteristic record rank
$k^*$ the approach to failure results in an acceleration of record breaking with 
decreasing
waiting times and increasing relative increments.
In our study the disorder of materials was represented by uniformly distributed 
failure thresholds.
We repeated the complete RB analysis for Weibull distributed disorder varying 
the Weibull
exponent. It has to be emphasized that the qualitative behavior of the results 
and the value of 
the exponents of the record size, increment and waiting 
time distributions all proved to be universal, only the characteristic load 
$\sigma_s^*$, furthermore, 
the scaling and cutoff exponents $\alpha$ and $\beta$ depend on the disorder.

Recently, the record breaking statistics of driven threshold models of complex 
systems
has been analyzed in  \cite{PhysRevE.87.052811}. These are cellular automata 
models of
self organized criticality where the slow external driving leads to the 
emergence of 
a steady state characterized by intermittent avalanches of relaxation events. 
Comparing the statistics of records of the models to the corresponding results
of IIDs the authors could point out correlations in the complex spatio-temporal 
evolution of avalanches. The most prominent deviation from IIDs was found for 
the 
Olami-Feder-Christensen model \cite{PhysRevLett.68.1244} where the number of 
records 
proved to increase as a power of the logarithm of the avalanche
number. The main difference of the dynamics of our FBM
and the above models is that during creep both the size and the rate of 
avalanches increase
so that no steady state arises. Hence, the comparison to IIDs in our case helped
to determine regimes controlled either by disorder or by the increasing stress 
level of
intact fibers. 
Record breaking statistics of inter-event times 
in aftershock sequences of earthquakes has recently been studied in 
\cite{npg-17-169-2010}.
Based on the non-homogeneous Poissonian process describing the rate of events, a 
power 
law behavior of the record number was obtained following the logarithmic 
increase. 
Our study demonstrates that the investigation of the record breaking statistics 
of the time 
series of crackling events reveals also interesting novel aspects of rupture 
phenomena.

\section*{Acknowledgment}
This research was supported by the European Union and the State of Hungary, 
co-financed by the European Social Fund in the framework of T\'AMOP 4.2.4. 
A/2-11-1-2012-0001 
``National Excellence Program''.
This work was supported by the European Commissions by the
Complexity-NET pilot project LOCAT and by NF\"U under the contract 
ERANET\_HU\_09-1-2011-0002. 
We also acknowledge the projects TAMOP-4.2.2.A-11/1/KONV-2012-0036 and OTKA 
K84157.

\bibliographystyle{frontiersinSCNS&ENG} 
\bibliography{/home/feri/papers/statphys_fracture}

\end{document}